\documentclass[conference]{IEEEtran}

\usepackage{amsmath}
\usepackage{amsfonts}
\usepackage{amssymb}
\usepackage{amsthm}
\usepackage{tabularx}
\usepackage{mathrsfs} 

\usepackage{url}
\usepackage{hyperref}

\newtheorem{remark}{Remark}

\usepackage{stfloats}
\usepackage{float}
\usepackage{graphicx}
\usepackage{cite}
\usepackage{xcolor}
\usepackage{subfigure}

\makeatletter
\def\blfootnote{\xdef\@thefnmark{}\@footnotetext}
\makeatother

\begin{document}
	
		\title{\huge{On Performance of FAS-aided Covert Communications}} 
\author{
	\IEEEauthorblockN{Farshad Rostami Ghadi\textsuperscript{†}, Masoud Kaveh\textsuperscript{‡}, Riku Jantti\textsuperscript{‡}, and F. Javier Lopez-Martinez\textsuperscript{†}}
	\IEEEauthorblockA{\textsuperscript{†}%Department of Electronic and Electrical Engineering, University College London, London, United Kingdom. 
		Department of Signal Theory, Networking and Communications, University of Granada, Granada, Spain.\\}
	\IEEEauthorblockA{\textsuperscript{‡}Department of Information and Communication Engineering, Aalto University, Espoo, Finland. \\}
	%\IEEEauthorblockA{\textsuperscript{§}Department of Signal Theory, Networking and Communications, University of Granada, Granada, Spain. \\}
	Emails: $\rm \{f.rostami, fjlm\}@ugr.es$; $\rm \{masoud.kaveh, riku.jantti\}@aalto.fi$.
}
	\maketitle
	\begin{abstract}
This paper investigates the impact of deploying the fluid antenna system (FAS) on the performance of covert communications. In particular, we focus on a scenario where a transmitter seeks to covertly communicate with a receiver, while a warden attempts to detect the transmission. Both the receiver and the warden are assumed to utilize planar FAS. We derive compact analytical expressions for the covertness outage probability (COP), defined as the complement of the sum of false alarm (FA) and missed detection (MD) probabilities. %Additionally, we present the outage probability (OP) of the legitimate transmission. 
By determining the optimal detection threshold that maximizes the COP, we characterize the success probability for the legitimate transmission, highlighting the trade-off between covertness and transmission success. Our numerical results confirm that while deploying FAS at the warden enhances its detection ability compared to fixed-position antennas (FPAs), equipping the receiver with FAS rather than FPAs significantly improves reception quality, leading to more reliable transmission.
	\end{abstract}
	%\vspace{-3.5ex}
	\maketitle
	%\blfootnote{\noindent Copyright (c) 2015 IEEE. Personal use of this material is permitted. However, permission to use this material for any other purposes must be obtained from the IEEE by sending a request to pubs-permissions@ieee.org.} 
%	\blfootnote{Manuscript received January 25, 2021; revised XXX. The review of this paper was coordinated by XXXX.} 
%		\blfootnote{This work has been funded in ..}
%	 	\blfootnote{\noindent The authors are with the .. (e-mail: $\rm$.}
	
%	\blfootnote{Digital Object Identifier 10.1109/XXX.2021.XXXXXXX}
	%\IEEEpeerreviewmaketitle
	\vspace{0mm}
	\section{Introduction}\label{sec-intro}
The fluid antenna system (FAS) is a promising technology that enhances wireless communication by dynamically adjusting antenna positions within a confined spatial region. Unlike traditional fixed-position antennas (FPAs), FAS offers adaptive port selection and improved spatial diversity, significantly boosting performance in challenging environments. This flexibility enables superior spectral efficiency, energy efficiency, and resilience to interference, positioning FAS as a critical enabler for the high-capacity, low-latency requirements of next-generation networks (NGNs), a.k.a., sixth-generation (6G) \cite{wong2020fluid,wong2020performance}.
Furthermore, the adaptability of FAS makes it highly compatible with emerging wireless paradigms, such as integrated sensing and communication (ISAC) \cite{zhou2024isac, wang2024isacdr}, reconfigurable intelligent surfaces (RIS) \cite{ghadi2024risfas, chen2024risfas}, and low-power Internet of Things (IoT) \cite{ghadi2024backscatter}, highlighting it as a key enabler for the next generation of intelligent, sustainable, and high-efficiency wireless systems.

However, as fluid antenna-enabled systems become more prevalent in 6G mobile technology, they face new security challenges, particularly from adversaries exploiting the system's spatial correlation to enhance eavesdropping capabilities \cite{new2024tutorial}. To mitigate such threats, physical layer security (PLS) strategies, such as artificial noise (AN) and power control (PC), have been integrated into FAS, demonstrating promising secrecy performance \cite{tang2023fas,ghadi2024secure,vega2024sop}. Despite these advancements, PLS does not address the critical challenge of making the communication undetectable to adversaries. Covert communication, where the goal is to hide the very existence of the transmission \cite{Lee2015}, is becoming increasingly vital, particularly in dense, interference-prone environments such as vehicle-to-vehicle (V2V) communications, where conventional methods may fail to prevent detection. 

Despite the growing interest in secure and covert communication, the role of FAS in covert systems remains largely unexplored. The only study considering FAS in covert communications was reported in \cite{Yao2024covert}, where only the transmitter (Alice) was equipped with FAS, while other nodes used FPAs. Specifically, \cite{Yao2024covert} optimized the secrecy rate of the receiver under a covertness constraint by jointly designing transmit beamforming and antenna positioning at the Alice, demonstrating that FAS significantly enhances both secrecy and covertness compared to FPA-based schemes. However, the impact of FAS on detection and transmission probabilities, particularly when both the receiver (Bob) and the warden (Willie) employ FAS, has not yet been investigated to the best of the authors' knowledge.

Motivated by this gap, we theoretically analyze the performance of FAS-aided covert communications. Specifically, we develop a compact analytical framework for evaluating the covertness outage probability (COP) and outage probability (OP) of transmission, incorporating PC strategies to assess the impact of FAS on covert communication. Furthermore, by determining the optimal detection threshold that maximizes COP\footnote{The optimal detection threshold is chosen to maximize the COP (i.e., maximize the probability that a transmission outage is declared because of covertness). This corresponds to the most favorable scenario for the warden in detecting the transmission and, conversely, the least favorable scenario for the transmitter Alice in maintaining covertness. This represents a worst-case analysis for covert communication, ensuring that the insights derived remain valid under the most stringent detection conditions \cite{shahzad2017covert,kim2022covert}.}, we characterize the success probability in terms of COP and OP, providing key insights into the fundamental trade-off between covertness and transmission reliability. Our findings reveal that while deploying FAS at the warden enhances its detection capability, equipping the receiver with FAS significantly improves reception quality, leading to more reliable transmission.
	\section{System Model}\label{sec-sys}
	\subsection{Channel Model}
\begin{figure}[!t]
	\centering
\includegraphics[width=0.85\columnwidth]{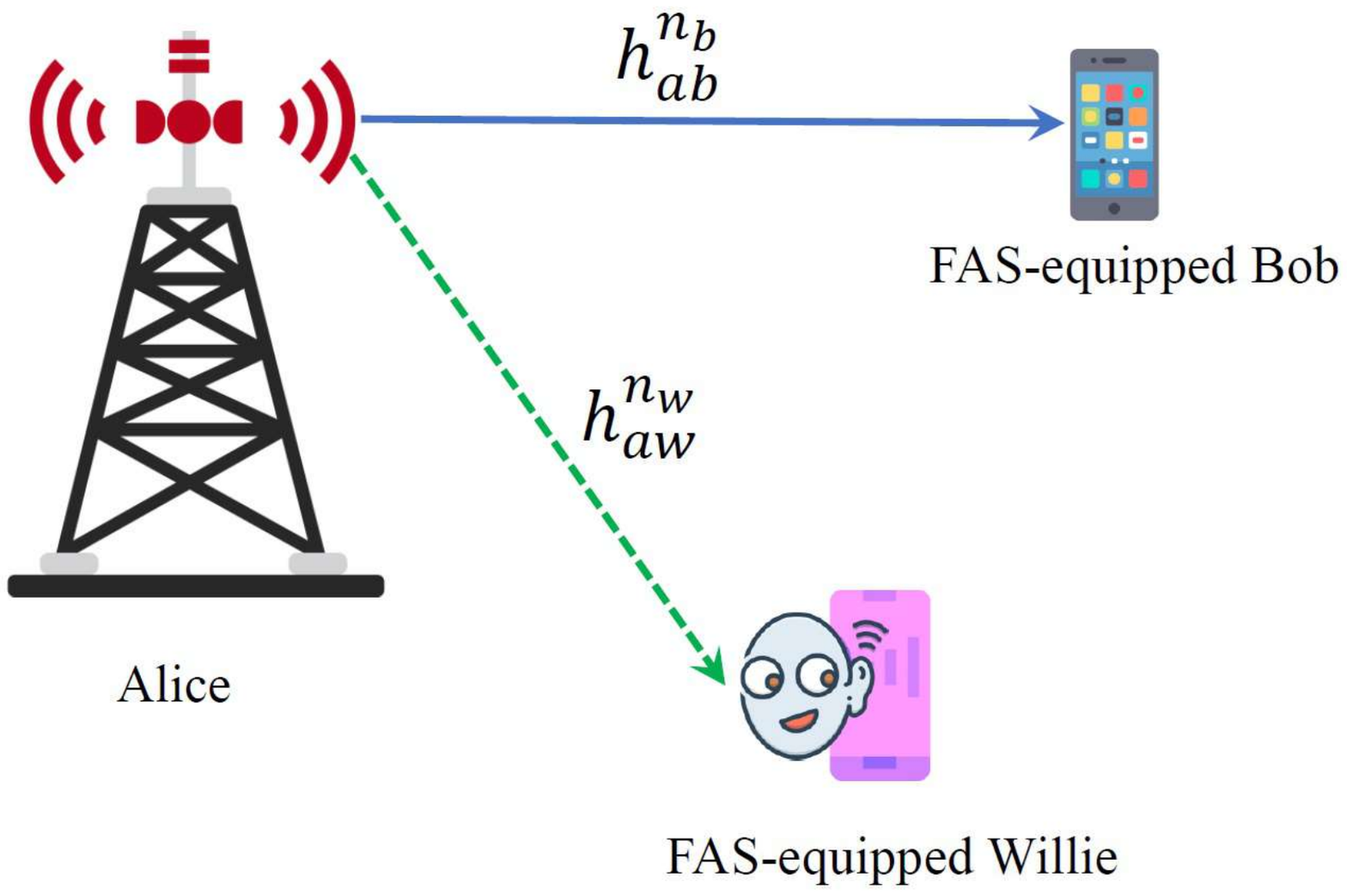}
	\caption{FAS-aided Covert Communications.}\vspace{0cm}
	\label{fig-model}
\end{figure}

	We consider a covert communication system as illustrated in \ref{fig-model}, where Alice aims to covertly communicate to Bob while Willie is attempting to detect the existence of this transmission. For notation simplicity, we use $a$, $b$, and $w$ define Alice, Bob, and Willie, respectively. %It is assumed that Alice and Bob perform in half-duplex mode whereas Willie can operate in full-duplex mode. 
     We assume that Alice has a single FPA, while both Bob and Willie are equipped by a planar FAS with $N_j=N_j^1\times N_j^2$ ports and size $W_j=\left(W_j^1\times W_j^2\right)\lambda^2$ for $j\in\left\{b,w\right\}$, where $\lambda$ denotes the wavelength. Moreover, a mapping function $\mathcal{F}\left(n_j\right)=\left(n_j^1,n_j^2\right)$ and its converse $n_j=\mathcal{F}^{-1}\left(n_j^1,n_j^2\right)$ are defined to transfer two-dimensional (2D) to one-dimensional (1D) index, in which $n_j\in\left\{1,\dots,N_j\right\}$ and $n_j^l\in\left\{1,\dots,N_j^l\right\}$ for $l\in\left\{1,2\right\}$. We also define the channel coefficient from $a$ to $n_j$-th port of node $j$ as $h_{aj}^{n_j}$, which is characterized as Rayleigh fading model. Hence, the corresponding channel gain $g_{aj}^{n_j}=\left|h_{aj}^{n_j}\right|^2$ follow the exponential distribution with unit mean. Moreover, it is assumed that Alice and Bob know the instantaneous and statistical channel state information (CSI) of $h_{ab}^{n_b}$, whereas Alice and Willie only know the statistical CSI of $h_{aw}^{n_w}$. Since the fluid antenna ports can freely switch to any position and be arbitrarily close to each other, the channel coefficients are spatially correlated. More specifically, the covariance between two arbitrary ports $n_j$ and $\tilde{n}_j$ at node $j$ under rich scattering is given by \cite{new2024information}
	\begin{align}
		\varrho_{n_j,\tilde{n}_j}=\mathcal{J}_0\left(2\pi\sqrt{\left(\frac{n_j^1-\tilde{n}_j^1}{N^1_j-1}W_j^1\right)^2+\left(\frac{n_j^2-\tilde{n}_j^2}{N^2_j-1}W_j^2\right)^2}\right),
	\end{align}
	where $\mathcal{J}_0\left(\cdot\right)$ defines the zero-order spherical Bessel function of the first kind.
	\subsection{Covert Transmission Scheme and Attaching Model}
We consider that Alice uses the PC transmission scheme, where she adjusts her transmit power $P_a$ to blend the message signals with background noise for covertness. 
%	Furthermore, we consider that Alice uses two transmission schemes, namely, power control (PC) and artificial noise (AN). In the PC scheme, Alice adjusts her transmit power $P_a$ to blend the message signals with background noise for covertness. In the AN scheme, Alice injects artificial noise into the message signals to confuse Willie, reducing his ability to detect the transmission. Unlike the PC scheme, the AN scheme uses a fixed transmit power $P_a$, splitting it between the message and noise, with $\beta\in\left(0,1\right]$, representing the fraction of power dedicated to the message.
	%\subsection{Attack Model}
	To detect Alice’s transmissions, Willie uses a standard likelihood ratio test (LRT) \cite{sobers2017covert}. He first sets a threshold $\zeta$ and measures the average power $\overline{P}_w$ of the symbols received during the transmission slot. If $\overline{P}_w \geq \zeta$, Willie concludes that Alice transmitted messages, defined as hypothesis $\mathcal{H}_1$. If $\overline{P}_w \leq \zeta$, he assumes no transmission occurred, denoted as hypothesis $\mathcal{H}_0$. Mathematically, the likelihood ratio test can be  expressed as 
    \begin{align}
	\overline{P}_w   \underset{\mathcal{H}_0}{\overset{\mathcal{H}_1}{\gtrless}}\zeta.
    \end{align}
	This test results in two types of errors: (i) false alarm (FA), where Willie mistakenly detects a transmission that did not happen; and (ii) missed detection (MD), where he fails to detect an actual transmission. We define the probabilities of these errors as $P_\mathrm{FA}$ and $P_\mathrm{MD}$, respectively. If neither error occurs, the transmission is considered to have maintained covertness, meaning Willie could not detect it. 
	\subsection{Signal Model}
When Alice transmits, she sends a vector of $k$ symbols, denoted by $\textbf{x}$, where each symbol $ \textbf{x}\left[i\right]$ for $i=1,\dots,k$ satisfies a unit power constraint such that $\mathbb{E}\left\{|x[i]|^2\right\} = 1 $. On the other hand, when Alice does not transmit, Willie only receives noise. Therefore, the received signal at $n_w$-th port of Willie is then expressed as
	\begin{align}\label{eq-yw}
	y^{n_w}_w = 
	\begin{cases} 
		\sqrt{P_a} h_{aw}^{n_w} \mathbf{x} + \mathbf{n}_w, &  \mathcal{H}_1, \\
		 \mathbf{n}_w  & \mathcal{H}_0,
	\end{cases}
	\end{align}
where $\mathbf{n}_w$ modeled as complex Gaussian noise with zero mean and variance $\sigma_w^2$, i.e., $\mathbf{n}_w\left[i\right]=\mathcal{CN}\left(0,\sigma_w^2\right)$.
	
To detect Alice's signal, Willie relies on the average received power $\overline{P}_w$, which given by
	\begin{align}\label{eq-pw}
	\overline{P}_w= \frac{1}{k} \sum_{i=1}^{k} \left|y_w^{n_j}\left[i\right]\right|^2=	\begin{cases}
			P_ag_{aw}^{n_w}+\sigma^2_w, &  \mathcal{H}_1,\\
			\sigma^2_w, & \mathcal{H}_0.
			\end{cases}
	\end{align}
Equation \eqref{eq-pw} is derived by applying the Strong Law of Large Numbers \cite{Browder2012math}
to \eqref{eq-yw} as the number of symbols grows large. This allows Willie to estimate the power of the received signal, which is critical for detecting whether Alice is transmitting, based on the threshold power.
Following the concept of FAS, we assume that only the best port $n_j^*$, which maximizes the received signal-to-noise ratio (SNR) at node $j$, is activated, i.e., 
\begin{align}
n_j^*=\underset{n}{\arg\max}\left\{\left|h_j^{n_{aj}}\right|^2\right\}
\end{align}
 Thus, the equivalent channel gain at node $j$ is defined as
\begin{align}
g_{\mathrm{fas},j}=\max\left\{g_{aj}^1,\dots,g_{aj}^{N_j}\right\}=g_{aj}^{n_j^*},
\end{align}
where $g_{aj}^{n_j^*}=\left|h_{aj}^{n_j^*}\right|^2$. Therefore, \eqref{eq-pw} can be rewritten as
	\begin{align}\label{eq}
	\overline{P}_w=	\begin{cases}
		P_ag_{\mathrm{fas},w}+\sigma^2_w, &  \mathcal{H}_1,\\
		\sigma^2_w, & \mathcal{H}_0.
	\end{cases}
\end{align}
	
 \section{Performance Analysis}
 
In this section, we derive analytical expressions for the COP, OP, and success probability to evaluate the performance of covert communication. 
	\subsection{COP}
	The COP is defined as the probability that Willie's operation is successful, i.e., the probability that neither a false alarm nor a missed detection happens \cite{Lee2015}, i.e., 
\begin{align}\label{eq-co}
	P_\mathrm{co} = 1-\left(P_\mathrm{FA} + P_\mathrm{MD}\right).
\end{align}
	If $\overline{P}_w\leq \zeta$, Willie concludes that Alice has not transmitted any messages and, as a result, fails to detect her transmission, leading to a MD event. Consequently, the probability of missed detection is denoted as
\begin{align}\label{eq-md}
	P_\mathrm{MD} &= \Pr \left( P_a g_{\mathrm{fas},w} + \sigma_w^2 \leq \zeta \right)\\
	& =\Pr\left(g_{\mathrm{fas},w}\leq\frac{\zeta-\sigma^2_w}{P_a}\right)\\
	&=\Pr\left(\max\left\{g_{aw}^1,\dots,g_{aw}^{N_w}\right\}\leq \frac{\zeta-\sigma^2_w}{P_a}\right)\\
	&\underset{=}{(a)}
	\begin{cases}
		C_w, & \zeta>\sigma^2_w,\\
		0, & \zeta\leq \sigma^2_w,
		\end{cases}
\end{align}
where \begin{align}\nonumber
C_w=T_{\nu_w,\mathbf{\Sigma}_w} \hspace{-2pt}\left(t_{\nu_w}^{-1} \hspace{-2pt}\left(1-\mathrm{e}^{-\frac{\zeta-\sigma^2_w}{P_a}}\right) \hspace{-2pt},\dots,t_{\nu_w}^{-1} \hspace{-2pt}\left(1-\mathrm{e}^{-\frac{\zeta-\sigma^2_w}{P_a}}\right) \hspace{-2pt}\right),
\end{align}
is the $t$-student copula, $t_{\nu_w}^{-1}\left(\cdot\right)$ is the inverse CDF (quantile function) of the univariate $t$-distribution having $\nu_w$ degrees of freedom,  $T_{\nu_w,\mathbf{\Sigma}_w}\left(\cdot\right)$ represents the CDF of the multivariate $t$-distribution with correlation matrix $\mathbf{\Sigma}_w$ and $\nu_w$ degrees of freedom. In fact, $(a)$ follows from approximating the correlation coefficient of Jakes' model using an elliptical copula, such as the $t$-student copula. Specifically, $\varrho_{n_j,\tilde{n}_j}\approx \Theta_{n_j,\tilde{n}_j}$, where $\Theta_{n_j,\tilde{n}_j}$ is the dependence parameter of the $t$-student copula, capturing the correlation between the corresponding ports \cite{ghadi2024g,ghadi2024rsma}.

If $\overline{P}_w\ge \zeta$, Willie concludes that Alice has transmitted a message, resulting in a FA mode, where he mistakenly detects a transmission when none occurred. Assuming that Willie is able to collect a sufficiently large number of samples, the probability of FA is given by 
\begin{align}\label{eq-fa}
P_\mathrm{FA} = \Pr \left( \sigma_w^2 \geq \zeta \right)=
\begin{cases} 
	0, &  \zeta > \sigma_w^2, \\
	1, &  \zeta \leq \sigma_w^2.
\end{cases}
\end{align}
Now, by applying \eqref{eq-md} and \eqref{eq-fa} into \eqref{eq-co}, COP is obtained as
\begin{align}
	P_{\mathrm{co}}=\begin{cases}
		1-C_w, & \zeta>\sigma^2_w,\\
		0, & \zeta\leq \sigma^2_w.
	\end{cases}
\end{align}
\begin{remark}
To maximize the COP, Willie needs to find the optimal threshold $\zeta^*$ that maximizes $P_\mathrm{co}$. Since $P_{\mathrm{co}} = 0$ when $\zeta \leq \sigma^2_w$, the optimal threshold must be in the region $\zeta > \sigma^2_w$. From the given expression of $C_w$, we see that $P_{\mathrm{co}}$ is a decreasing function of $\zeta$, as $C_w$ increases with $\zeta$. This means that to maximize $P_{\mathrm{co}}$, $C_w$ should be  minimized, which occurs when $\zeta$ is as small as possible while still satisfying $\zeta>\sigma^2_w$. Thus, the optimal threshold is $
\zeta^* = \sigma^2_w + \mu$ for an arbitrarily small $\mu > 0$. 
\end{remark}
\subsection{OP}
In the context of our covert communication model, the OP is a critical performance metric that reflects the likelihood of Alice and Bob's communication being unsuccessful due to insufficient channel capacity. Specifically, the OP is defined as the probability that channel capacity between Alice and Bob $\mathcal{C}_b$ is less than a desired target rate $R_b$, i.e., 
\begin{align}
	&P_\mathrm{out} = \Pr\left(\log_2\left(1+\frac{P_ag_{\mathrm{fas},b}}{\sigma^2}\right)\leq R_b\right)\\
	&=\Pr\left(\max\left\{g_{ab}^1,\dots,g_{ab}^{N_b}\right\}\leq \overline{R}_b\right)\\
	&=T_{\nu_b,\mathbf{\Sigma}_b}\left(t_{\nu_b}^{-1}\left(1-\mathrm{e}^{-\overline{R}_b}\right),\dots,t_{\nu_b}^{-1}\left(1-\mathrm{e}^{-\overline{R}_b}\right)\right)
\end{align}
where $\overline{R}_b=\left(2^{R_b}-1\right)\sigma_b^2/P_a$.

\subsection{Success Probability}
In covert communication, ensuring a successful transmission while maintaining covertness presents a fundamental trade-off. The probability of a successful transmission depends on two key factors: the ability to avoid detection by the warden and the reliability of the communication link between Alice and Bob. Alice's transmission power is carefully chosen so that transmission is not detected by Willie while Bob being able to decode the message. Willie also needs to design the threshold to avoid false alarm events, but at the same time not missing the detection of a transmission. To quantify the covertness-reliability trade-off, we define the success probability as the likelihood that Alice's transmission remains covert while also meeting the minimum required communication rate at Bob. We define the success probability to reflect only cases where a message is actually transmitted. Specifically, success probability is defined as the joint probability of Alice's message being reliably received at Bob while remaining undetected at Willie, i.e., 
	\begin{align}\label{eq-suc}
		P_\mathrm{suc}=P_\mathrm{MD}(\zeta^*)\times(1-P_\mathrm{out}),
		\end{align}		
where $P_\mathrm{MD}(\zeta^*)$ represents the probability of MD at Willie given the optimal detection threshold\footnote{While we explicitly exclude $P_{\rm FA}$ from the success probability definition, we note that the threshold $\zeta^*$ at Willie is designed by taking $P_{\rm FA}$ into consideration.} $\zeta^*$, ensuring that Alice's transmission remains covert. 
\begin{figure}[!t]
	\centering
	\includegraphics[width=0.9\columnwidth]{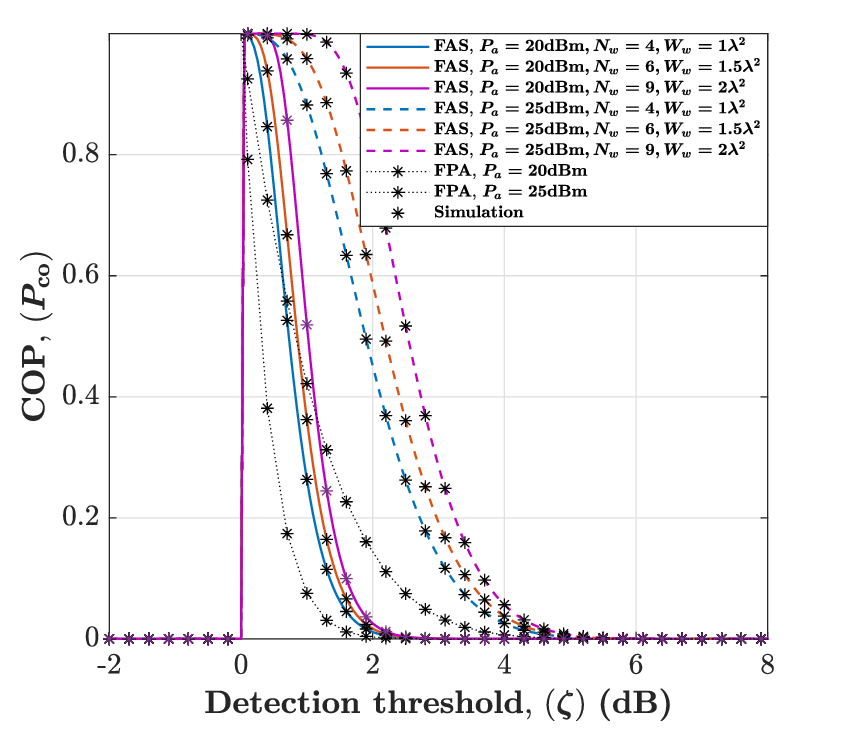}
	\caption{COP versus detection threshold $\zeta$ for selected values of $P_a$, $N_w$, and $W_w$.}\vspace{0cm}
	\label{fig_cop_zeta}
\end{figure}
\begin{figure}[!t]
	\centering
	\includegraphics[width=0.9\columnwidth]{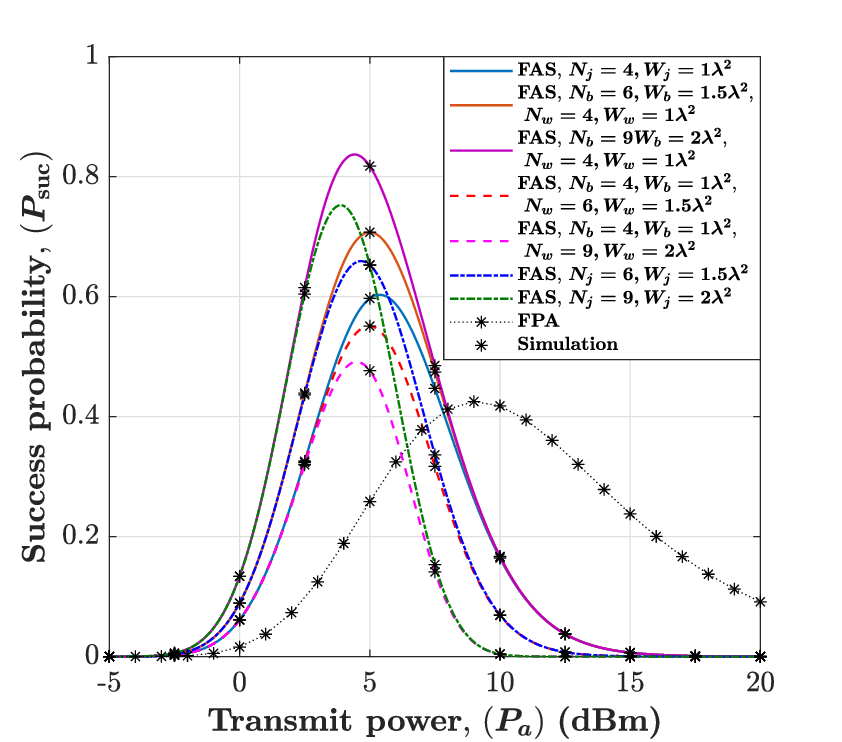}
	\caption{Success probability versus transmit power $P_a$ three different scenarios.}\vspace{0cm}
	\label{fig_suc_p}
\end{figure}
\section{Numerical Results}\label{sec-num}
Here, we provide the numerical results to assess the performance of the proposed FAS-aided covert communication setup. In this regard, we set the parameters as $\mu=0.01$, $\nu_j=40$, $\sigma^2_w=0$ dB, $\sigma^2_b=-20$ dB, $R_b=0.5$ bits, $N_j=4$, $W_j=1\lambda^2$, and  $P_a\in\left\{20, 25\right\}$ dBm. We also consider the FPA-based scheme as the benchmark. 

Fig. \ref{fig_cop_zeta} illustrates the behaviour of the COP in terms of the detection threshold $\zeta$ for various scenarios. As expected, aligning with Remark 1, COP is maximized when $\zeta$ is just slightly above $\sigma_w^2$, confirming that Willie should set $\zeta^*=\sigma^2_w+\mu$ to achieve the most favorable detection performance\footnote{In practice, the warden has uncertainty about its own receiver’s noise power \cite{Lee2015}, which may affect his detection capabilities and hence be beneficial to achieve covertness.}. We can see that for small values of $\zeta$, COP remains close to 1, indicating that the transmission is highly detectable. As $\zeta$ increases, COP rapidly decreases, showing that Willie's detection capability deteriorates at higher thresholds, which is reasonable since a larger detection threshold means that Willie becomes more conservative in declaring the presence of a transmission. This reduces the probability of FA but increases the likelihood of MD, making it harder for Willie to reliably detect Alice's transmission. Essentially, as $\zeta$ grows, the received signal at Willie is less likely to exceed the threshold, leading to more frequent missed detections and a lower COP. Additionally, it is observed that deploying FAS enhances Willie's detection ability compared with FPA-enabled model, leading to a steeper decline in COP, and thus, gives Willie a significant advantage in detecting the covert transmission. This improvement is mainly because the FAS allows for dynamic positioning, optimizing signal reception and adaptability compared to the FPAs.  Moreover, when Alice increases her transmit power from $20$ dBm to $25$ dBm, COP decreases for a given threshold. This aligns with the fact that a higher $P_a$ makes the signal more detectable, which is beneficial for Willie's detection performance and compromises covertness. The difference between the $20$ dBm and $25$ dBm cases is more pronounced for FAS than for FPA, showing that FAS amplifies Willie's detection improvements.

Fig. \ref{fig_suc_p} illustrates the relationship between the success probability and the transmit power $P_a$, under different configurations of the FAS at Bob and Willie. At low $P_a$, Bob's OP is high, leading to poor reception despite lower detection at Willie. As increases $P_a$, $P_\mathrm{suc}$ reaches an optimal point where Bob's reception improves while Willie still struggles with detection. However, beyond this peak, further increasing $P_a$ makes it easier for Willie to detect Alice's transmission, reducing $P_\mathrm{suc}$. Additionally, the figure considers three distinct FAS configurations: (i) increasing Bob's number of ports $N_b$ and antenna size $W_b$ while keeping Willie's fixed, (ii) increasing Willie’s number of ports $N_w$ and antenna size $W_w$ while keeping Bob's fixed, and (iii) simultaneously increasing FAS parameters for both. When $N_b$ and $W_b$ are increased while keeping Willie's FAS parameters constant, a notable improvement in $P_\mathrm{suc}$ is observed. This occurs because increasing $N_b$ and $W_b$ enhances the spatial diversity at Bob, effectively mitigating fading effects and increasing the received SNR. As a result, $P_\mathrm{out}$ is significantly reduced, ensuring more reliable reception. Furthermore, since Willie's detection capabilities remain unchanged in this scenario, the $P_\mathrm{MD}$ does not deteriorate. This combination, i.e., better reception at Bob without an increase in Willie's detection ability, maximizes $P_\mathrm{suc}$.

In contrast, when $N_w$ and $W_w$ are increased while keeping Bob's parameters unchanged, then $P_\mathrm{suc}$ decreases. This behavior arises because increasing $N_w$ and $W_w$ grants Willie better spatial adaptability, allowing him to exploit favorable fading conditions to detect Alice's transmission more accurately. Consequently, $P_\mathrm{MD}$ decreases, reducing $P_\mathrm{suc}$. However, the reduction in $ P_{\mathrm{suc}}$ is not as severe as the improvement observed when enhancing Bob’s FAS due to the asymmetry between reception (coherent) and detection (energy-based). Bob needs a strong SNR to decode messages, so increasing $N_b$ and $W_b$ directly reduces $P_\mathrm{out}$. In contrast, Willie's detection relies on hypothesis testing, where increasing $N_w$ and $W_w$ improves detection but with a different scaling law. 

When the number of ports and antenna size are increased at both Bob and Willie, an overall enhancement in $P_\mathrm{suc}$ is observed, albeit with some trade-offs. On the one hand, Bob’s improved spatial diversity significantly reduces $P_\mathrm{out}$, improving the reliability of message reception. On the other hand, Willie also benefits from enhanced spatial adaptability, which increases his detection probability and reduces $P_\mathrm{MD}$. Interestingly, the net result still favors an increase in $P_\mathrm{suc}$. This suggests that Bob's ability to exploit spatial diversity for reliable reception outweighs Willie’s detection gain, likely due to Bob’s advantage in processing power and the fact that covert communication relies on both noise uncertainty and channel variations. However, compared to the case where only Bob’s FAS is improved, the peak value of $P_\mathrm{suc}$ in this scenario is slightly lower, as Willie's enhanced detection capabilities counteract some of the improvements at Bob.

%Fig. \ref{fig_suc_p} illustrates the success probability as a function of the transmit power $P_a$, demonstrating the fundamental trade-off between covertness and transmission success. As expressed in \eqref{eq-suc}, $P_\mathrm{suc}$ is defined at the optimum detection scenario for Willie. This figure reveals that increasing $P_a$ significantly enhances success probability across all configurations. This behavior is expected since higher transmit power improves the received signal strength at Bob, thereby reducing $P_\mathrm{out}$ and increasing the likelihood of successful transmission. However, excessive transmit power also increases the risk of detection by Willie, potentially lowering $P_\mathrm{co}\left(\zeta^*\right)$, thereby introducing a trade-off between reliability and covertness. Comparing the different schemes, it is evident that FAS achieves a notably higher success probability than FPA, as indicated by the dotted black curve. The FAS configuration benefits from enhanced spatial diversity, allowing for more favorable channel conditions and a lower probability of transmission failure. Furthermore, within the FAS schemes, increasing the number of antennas $N_b$ and the spatial region at Bob $W_b$ leads to further improvements in $P_\mathrm{suc}$. Notably, increasing $N_w$ and $W_w$ at Willie enhances his detection capability but has a negligible impact on $P_\mathrm{suc}$, as Bob's success probability is primarily governed by his own antenna configuration and received signal quality rather than Willie’s ability to detect the transmission.
\section{Conclusion}\label{sec-con}
This study explored the role of FAS in covert communication scenarios, where a transmitter aimed to communicate undetected with a receiver while a warden attempted to intercept the transmission. By incorporating planar FAS at both the receiver and the warden, analytical expressions were derived for COP, accounting for FA and MD probabilities, as well as OP to assess transmission reliability. The optimal detection threshold that maximized COP was identified, offering insights into the fundamental trade-off between covertness and successful message delivery. The results demonstrated that while deploying FAS at the warden enhanced its detection capability relative to FPAs, integrating FAS at the receiver significantly improved signal reception, ensuring a more reliable communication link.

\section*{Acknowledgment}
{This work is supported in part by the European Union's Horizon 2022 Research and Innovation Programme under Marie Skłodowska-Curie Grant No. 101107993, and in part by grant PID2023-149975OB-I00 (COSTUME) funded by MICIU/AEI/10.13039/501100011033.} 
%\bibliographystyle{IEEEtran}
%\bibliography{refs.bib}

\begin{thebibliography}{1}
	
	\bibitem{wong2020fluid}  
	K. K. Wong, A. Shojaeifard, K. F. Tong, and Y. Zhang, “Fluid antenna systems,” \textit{IEEE Trans. Wireless Commun.}, vol. 20, no. 3, pp. 1950–1962, Mar. 2020.
	\bibitem{wong2020performance}
	K. K. Wong, A. Shojaeifard, K. -F. Tong and Y. Zhang, "Performance Limits of Fluid Antenna Systems," \emph{IEEE Commun. Lett.}, vol. 24, no. 11, pp. 2469-2472, Nov. 2020.

\bibitem{zhou2024isac}  
L. Zhou, J. Yao, M. Jin, T. Wu, and K. K. Wong, “Fluid antenna-assisted ISAC systems,” \textit{IEEE Wireless Communications Letters}, vol. 13, no. 12, pp. 3533–3537, Dec. 2024.
\bibitem{wang2024isacdr}  
C. Wang, G. Li, H. Zhang, K. K. Wong, Z. Li, D. W. K. Ng, and C. B. Chae, “Fluid antenna system liberating multiuser MIMO for ISAC via deep reinforcement learning,” \textit{IEEE Transactions on Wireless Communications}, vol. 23, no. 9, pp. 10879–10894, Sep. 2024.

\bibitem{ghadi2024risfas}  
F. R. Ghadi, K. K. Wong, W. K. New, H. Xu, R. Murch, and Y. Zhang, “On performance of RIS-aided fluid antenna systems,” \textit{IEEE Wireless Communications Letters}, vol. 13, no. 8, pp. 2175–2179, Aug. 2024.
\bibitem{chen2024risfas}  
J. Chen, Y. Xiao, J. Zhu, Z. Peng, X. Lei, and P. Xiao, “Low-complexity beamforming design for RIS-assisted fluid antenna systems,” in \textit{Proc. IEEE Int. Conf. Commun. Workshops (ICC Workshops)}, Jun. 2024, pp. 1377–1382.

%\bibitem{ghadi2024fasnoma}  
%F. R. Ghadi, M. Kaveh, K. K. Wong, R. Jäntti, and Z. Yan, “On performance of FAS-aided wireless powered NOMA communication systems,” in \textit{Proc. 20th Int. Conf. Wireless Mobile Comput., Netw. Commun. (WiMob)}, Oct. 2024, pp. 496–501.
\bibitem{ghadi2024backscatter}  
F. R. Ghadi, M. Kaveh, K. K. Wong, and Y. Zhang, “Performance analysis of FAS-aided backscatter communications,” \textit{IEEE Wireless Communications Letters}, vol. 13, no. 9, pp. 2412–2416, Sep. 2024.

    
	\bibitem{new2024tutorial}  
	W. K. New \textit{et al.}, “A tutorial on fluid antenna system for 6G networks: Encompassing communication theory, optimization methods and hardware designs,” \emph{IEEE Commun. Surv. Tutor.}, 2024.
	\bibitem{tang2023fas}  
	B. Tang, H. Xu, K. K. Wong, K. F. Tong, Y. Zhang, and C. B. Chae, “Fluid antenna enabling secret communications,” \emph{IEEE Commun. Lett.}, vol. 27, no. 6, pp. 1491–1495, Jun. 2023.
	\bibitem{vega2024sop}  
	J. D. Vega-Sánchez, L. Urquiza-Aguiar, H. R. C. Mora, N. V. O. Garzón, and D. P. M. Osorio, “Fluid antenna system: Secrecy outage probability analysis,” \emph{IEEE Trans. Veh. Technol.}, vol. 73, no. 8, pp. 11458–11469, Aug. 2024.
	\bibitem{ghadi2024secure}  
	F. R. Ghadi \textit{et al.}, “Physical layer security over fluid antenna systems: Secrecy performance analysis,” \emph{IEEE Trans. Wireless Commun.}, vol. 23, no. 12, pp. 18201–18213, Dec. 2024. 
\bibitem{Lee2015}
S. Lee, R. J. Baxley, M. A. Weitnauer and B. Walkenhorst, "Achieving Undetectable Communication," \emph{IEEE J. Sel. Top. Signal Process.}, vol. 9, no. 7, pp. 1195-1205, Oct. 2015.
\bibitem{Yao2024covert}  
J. Yao \textit{et al.}, “FAS for secure and covert communications,” \emph{arXiv preprint}, \url{arXiv:2411.09235},  Nov. 2024.
\bibitem{shahzad2017covert}
K. Shahzad, X. Zhou and S. Yan, "Covert Communication in Fading Channels under Channel Uncertainty," \emph{2017 IEEE 85th Vehicular Technology Conference (VTC Spring)}, Sydney, NSW, Australia, 2017, pp. 1-5.
\bibitem{kim2022covert}
S. W. Kim and H. Q. Ta, "Covert Communications Over Multiple Overt Channels," \emph{IEEE Trans. Commun.}, vol. 70, no. 2, pp. 1112-1124, Feb. 2022.
\bibitem{new2024information}
W. K. New \textit{et al.}, "An Information-Theoretic Characterization of MIMO-FAS: Optimization, Diversity-Multiplexing Tradeoff and q-Outage Capacity," \emph{IEEE Trans. Wireless Commun.}, vol. 23, no. 6, pp. 5541-5556, June 2024.
\bibitem{sobers2017covert}
T. V. Sobers \textit{et al.}, "Covert Communication in the Presence of an Uninformed Jammer," \emph{IEEE Trans. Wireless Commun.}, vol. 16, no. 9, pp. 6193-6206, Sept. 2017.
\bibitem{Browder2012math}
A. Browder, \emph{Mathematical analysis: an introduction}. Springer Science
\& Business Media, 2012.
\bibitem{ghadi2024g}
F. R. Ghadi \textit{et al.}, "A Gaussian Copula Approach to the Performance Analysis of Fluid Antenna Systems," \emph{IEEE Trans. Wireless Commun.}, vol. 23, no. 11, pp. 17573-17585, Nov. 2024.
\bibitem{ghadi2024rsma}  
F. R. Ghadi \textit{et al.}, “Fluid Antenna-Aided Rate-Splitting Multiple Access,” \emph{arXiv preprint}, \url{arXiv:2411.11453},  Nov. 2024.
	\end{thebibliography}

\end{document}